# A Cautionary Note on the Zagarola and Smits Similarity Parameter for the Turbulent Boundary Layer

David Weyburne[1]
Air Force Research Laboratory
2241 Avionics Circle
WPAFB OH 45433

Zagarola and Smits (M. Zagarola and A. Smits, 1998, Mean-flow scaling of turbulent pipe flow. *J. Fluid Mech*. **373**, 33–79) developed an empirical velocity parameter for scaling the outer region of the turbulent boundary layer velocity profile that has been widely applied to experimental datasets. Plots of the scaled defect profiles indicate that most datasets display similar-like behavior using the Zagarola and Smits scaling parameter. In the work herein, it is shown that the common practice of finding similarity behavior using the defect profile is often incomplete in the sense that not all of the criteria for similarity have been checked for compliance. When full compliance is checked, it is found that most of the datasets which display defect similarity do not satisfy all the criteria required for similarity. The nature of this contradiction and noncompliance is described in detail. It is shown that the original datasets used by Zagarola and Smits display this flawed similarity behavior. Hence, a careful reassessment of any claims in the literature is required for those groups that attempted to use the defect profile and the Zagarola and Smits type of velocity scaling parameter to assert similarity of the velocity profile.

## 1. Introduction

In a now very influential set of papers, Zagarola and Smits [1,2] developed an empirical scaling parameter for the velocity profile in the outer region of the wall-bounded turbulent boundary layer. This velocity scaling parameter has been successful at producing similar behavior in defect velocity profiles taken at different stations in the flow direction for many turbulent boundary layer datasets. The two papers together have been cited in 600 research works. For pipe flow, Zagarola and Smits [1] and McKeon and Morrison [3] showed that this velocity scaling parameter collapses defect velocity profiles to a single profile for datasets taken at Princeton's Super Pipe facility. For turbulent boundary layers on a plate, the equivalent velocity scaling parameter has shown good success in collapsing profiles over a wide Reynolds number range [2]. Building on these results, Castillo and George [4], Castillo and Walker [5], and Cal and Castillo [6] (and references therein) have examined an extensive set of experimental datasets and claim that most turbulent boundary layer datasets appear to show similar behavior when scaled with the Zagarola and Smits velocity scale. Recent reviews on turbulent velocity profile scaling by Panton [7] and Buschmann and Gad-el-Hak [8] confirm Zagarola and Smits' success and thereby add support to Castillo and George's claim that most turbulent boundary layer datasets display similar behavior when scaled with the Zagarola and Smits scaling parameter.

However, recently it was reported by Weyburne [9] that the analysis of Castillo, George, and coworkers [4-6] was flawed due to an incomplete assessment of the compliance for similarity. Castillo, George, and coworkers' search for similarity is based on the flow governing equation approach to similarity developed by Castillo and George [4]. This theoretical approach lead to the development of a number of flow criteria necessary for a set of profiles to display similar behavior. Using one of the new constraints as a search criterion, in this case the pressure gradient parameter $\Lambda$ equal to a constant, Castillo, George, and coworkers found that most turbulent boundary layer experimental datasets exhibited velocity profile similarity when scaled with the Zagarola and Smits velocity scale. For similarity of the outer region of the velocity profiles, they based their claims of similarity on examination of plots of the defect profiles, defined as $u_e(x)-u(x,y)$ where $u(x,y)$ is the velocity in the flow direction (*x*-direction) and $u_e(x)$ is the corresponding velocity at the boundary layer edge. If the scaled profiles from station to station plot on top of one another then similarity is assumed. Scaled defect profile plots from a wide variety of datasets did indeed collapse toward a single curve indicative of similar behavior. However, Weyburne [9] showed that when crossed checked, the same data and scaling parameters that were replotted as scaled velocity profiles no longer showed similar behavior. This would seem to indicate that it is possible to change the physics of the flow merely by replotting the data using a simple shift of the profile data (subtracting the data by the endpoint). This paradoxical behavior lead to the discovery of what Weyburne termed "true similarity" and "false similarity." The false similarity case occurs when defect similarity is present but not all of the criteria for similarity are satisfied as required by the flow governing equation approaches to similarity developed by Rotta [10], Townsend [11], Castillo and George [4], and Jones, Nickels, and Marusic [12]. The exact nature of this false similarity problem is detailed below.

[1]Email: David.Weyburne@gmail.com

In what follows we begin by explaining the reason for the remarkable results obtained with the Zagarola and Smits scaling parameter. Next, we outline the nature of the defect profile similarity problem. We go on to show that this false similarity found by Weyburne [9] extends to the experimental pipe and plate flow data presented by Zagarola and Smits [1,2] and McKeon and Morrison [3].

## 2. Defect Profile and the Zagarola and Smits scaling parameter

Before we go into an explanation the paradoxical behavior, it is instructive to understand the reason for the apparent success of the Zagarola and Smits scaling parameter. Rather than working with the velocity profile directly, it has become common to discuss turbulent boundary layer similarity in terms of the defect profile. Adapting the standard definition of similarity [13], we can define defect profile similarity as the case when two profiles taken from different stations along the flow differ only by simple scaling parameters in $y$ and $u_e-u(x,y)$ where the *x*-direction is with the flow, *y*-direction is the direction perpendicular to the wall, $u(x,y)$ is the velocity in the flow direction, and $u_e$ is value of $u(x,y)$ at the boundary layer edge. If we take the length scaling parameter as $\delta_s$ and velocity scaling parameter as $u_s$, then defect profile similarity is defined as

$$\frac{u_e(x_1)-u\left(x_1,y/\delta_s(x_1)\right)}{u_s(x_1)} = \frac{u_e(x_2)-u\left(x_2,y/\delta_s(x_2)\right)}{u_s(x_2)} \quad \text{for all } y. \tag{1}$$

Experimentally, it is difficult to use Eq. 1 to test for similarity directly since the $y/\delta_s$ values are not the same from one measurement station to the next station as $\delta_s$ is a function of *x*. Hence the usual method to test for similarity is to plot all of the scaled defect profiles from various measurement stations along the flow on one graph. Similarity is indicated if all of the profiles plot on top of one another.

What makes the Zagarola and Smits scaling parameter so special in terms of defect profile similarity? The answer lies in the graphs of the scaled profiles. Consider the area under these scaled profiles plotted against the scaled wall height. Mathematically, the area under the scaled defect profile plotted against the scaled height is given by

$$a(x) \;=\; \int_0^{h/\delta_s} d\left\{\frac{y}{\delta_s}\right\} \frac{u_e - u(x, y/\delta_s)}{u_s} \quad , \tag{2}$$

where $h$ is located deep in the free stream. A necessary, but not sufficient, condition for similarity is that the areas must be equal, i.e. $a(x_1) = a(x_2)$. For flow along a wall the Zagarola and Smits velocity scaling parameter is defined as $u_{ZS} = u_e\delta_1/\delta$ where $\delta_1$ is the displacement thickness and $\delta$ is the boundary layer thickness. If we take $u_s = u_{ZS}$ and $\delta_s = \delta$, then Eq. 2 becomes

$$\begin{aligned} a(x) \;&=\; \int_0^{h/\delta} d\left\{\frac{y}{\delta}\right\} \frac{u_e - u(x, y/\delta)}{u_e\delta_1/\delta} \\ a(x) \;&=\; \frac{1}{\delta_1/\delta}\frac{1}{\delta}\int_0^{h} dy\, \frac{u_e - u(x,y)}{u_e} \\ a(x) \;&=\; \frac{1}{\delta_1}\int_0^{h} dy\, \frac{u_e - u(x,y)}{u_e} \\ a(x) \;&=\; 1 \quad . \end{aligned} \tag{3}$$

Hence the area under the scaled defect profile plotted versus the scaled height will always be normalized and be equal to one using $u_s = u_{ZS}$ and $\delta_s = \delta$. As long as the profile shape does not significantly change from one measurement station to another along the plate, then the scaled defect profile curves will tend to plot on top of one another since the areas are all normalized. This area normalization explains the remarkable success achieved using the defect profile and the Zagarola and Smits scaling. However, it is important to point out that having equal areas is a necessary but not a sufficient condition to insure similarity [14]. It is this critical fact that must be considered when using the Zagarola and Smits scaling parameter as we show below.

### 3. Defect Profile Similarity

So far it has not been demonstrated that there is any problem with the Zagarola and Smits scaling parameter. To understand where this false similarity behavior discussed by Weyburne [9] comes into play, we start by looking at the traditional definition of similarity of the velocity profile given by Schlichting [13] together with a simple similarity equivalency derivation used by Weyburne [14]. We have already defined defect profile similarity with Eq. 1.

Using the same notations, velocity profile similarity is defined [13] as the case where the scaled profile at a station $x_1$ and the scaled profile at $x_2$ satisfy the condition

$$\frac{u\left(x_1, y/\delta_s\left(x_1\right)\right)}{u_s\left(x_1\right)} = \frac{u\left(x_2, y/\delta_s\left(x_2\right)\right)}{u_s\left(x_2\right)} \quad \text{for all } y. \tag{4}$$

By inspection of Eqs. 1 and 4, one sees that the defect profile similarity will be equivalent to velocity profile similarity if

$$\frac{u_e\left(x_1\right)}{u_s\left(x_1\right)} = \frac{u_e\left(x_2\right)}{u_s\left(x_2\right)} \quad . \tag{5}$$

Eq. 5 is an important equation. Not only is it a requirement for equivalence of the defect and velocity profile similarity but this equation also shows up as a criterion for defect profile similarity in flow governing equation approaches to similarity as developed by Rotta [10] (see his Eq. 14.3), Townsend [11] (see his Eq. 7.2.3), Castillo and George [4] (see their Eq. 9), and Jones, Nickels, and Marusic [12] (see their Eq. 3.9, a2+a4). (Note that in Rotta's formulation, Rotta assumed $u_s(x)=u_\tau(x)$ but regardless, all four formulations have the identical criterion that $u_e/u_s$ must be the same constant value from one station to the next). Furthermore, in Appendix A we offer additional theoretical derivations that indicate that $u_e/u_s$ equal to a constant is required for similar behavior. This point is worth emphasizing; since similarity requires that $u_e/u_s$ be constant for the profiles to be similar, then this in turn requires that defect profile similarity must be accompanied by velocity profile similarity. You cannot have one without the other. The fact that Weyburne [9] found many instances in which datasets used by Castillo, George, and coworkers showed defect profile similarity but not velocity profile similarity is therefore both significant and at first perplexing.

Although the similarity requirement that $u_e/u_s$ be constant has been known for years, in actual practice it not usually tested to ensure that the condition is met. It is easy to understand why. In the past, the search for similarity in turbulent boundary layers datasets consists of simply plotting datasets using the various candidate $u_s$ and $\delta_s$ values. If the defect profile plots show similar behavior then the thinking was that there is no need to do additional checking. However, using side-by-side comparisons of the scaled defect profile and scaled velocity profile plots, Weyburne [9] showed that many of the datasets used by Castillo, George, and coworkers [4-6] to claim defect profile similarity do not satisfy the Eq. 5 criterion, that is, $u_e/u_s$ was not a constant. This points to a logical inconsistency: defect similarity is apparently present but one of the requirements for defect profile similarity is not satisfied. To understand how this can happen we observe that the defect profile in the outer tail region is being forced to zero so any differences in the $u_e/u_s$ values are not easily observed using the normal *x-y* plotting limits for defect profiles. Any differences in the $u_e/u_s$ values are in fact being hidden. Now, for the scaled velocity profile plots, one can verify whether or not $u_e/u_s$ is a constant by simply looking in the boundary layer edge region since these values will be $u_e/u_s$ values by definition. If the plotted scaled velocity profiles do not collapse to single curve in this tail region then this condition is not met and similarity is not present. Weyburne [9] showed that in eleven of the datasets used by Castillo, George, and coworkers to assert defect similarity, none of the datasets showed velocity profile similarity as evidenced in the boundary layer edge region. Hence, in those cases, the similarity condition $u_e/u_s$ be constant is NOT satisfied. This same behavior was also found in the datasets used in the scaling review papers by Panton [7] and Buschmann and Gad-el-Hak [8]. This behavior, showing defect profile similarity but not velocity profile similarity, is what Weyburne [9] termed false similarity.

In order to make this point more transparent, we will test some additional datasets below to see whether the similarity condition $u_e/u_s$ be constant is satisfied or not. We note that Zagarola and Smits [1,2] and McKeon and Morrison [3] used defect profile plots to attempt to claim similarity behavior. Checking these datasets will therefore let us answer the question whether the same problem found in Castillo, George, and coworkers work is also present in their data.

**3. Experimental Profile Check**

The above analysis points to the need to check for complete similarity by verifying Eq. 5 criterion compliance. One way to do this is to calculate the ratio $u_e(x)/u_s(x)$ and verify that it is a constant for the profiles in question. However, an easier method is to simply compare plots of the scaled defect profile side-by-side with plots of the scaled velocity profile. According to the analysis above, both should display similar behavior if Eq. 5 criterion is satisfied and similarity is present. Plots of the scaled velocity profiles can be viewed to verify that at the boundary layer edge the ratio $u_e(x)/u_s(x)$ is constant (or not), i.e. do the scaled profiles collapse to a single curve near the boundary layer edge. We use this side-by-side comparison to test some of the datasets Zagarola and Smits used to claim outer region similarity. We begin with some Super Pipe flow data plotted in Zagarola and Smits [1] as their Fig. 24b. We reproduce their Fig. 24b here as our Fig. 1a. In this figure, the length scaling parameter is the pipe radius *R* and the velocity scaling parameter is $u_{ZS} = u_{CL} - \bar{u}$ where $u_{CL}$ is the center line velocity and $\bar{u}$ is the average velocity. In Fig. 1b we plot the same velocity data using the *y*-axis scale $u/u_{ZS}$ instead of $(u_{CL} - u)/u_{ZS}$. It is apparent in this plot that the outer region of the boundary layer does not show similarity when the Zagarola and Smits scaling parameter is used to scale this velocity profile dataset.

Consider another example given by McKeon and Morrison [3] in which they claim that *R* and the velocity scale $u_{ZS} = u_{CL} - \bar{u}$ results in similarity collapse of the data from a more recent Super Pipe dataset. We reproduce part of their Fig. 2a here as our Fig. 2a. In Fig. 2b we plot the same velocity data using the *y*-axis scale $u/u_{ZS}$ instead of $(u_{CL} - u)/u_{ZS}$. Again, while the scaled defect profiles appear to be similar, the scaled velocity profile plots do not display similar behavior in the outer region near the boundary layer edge.

For flow data on a plate, Zagarola and Smits [2] used data from a number of sources including Smith [15]. For this dataset, they used the *y* length scale $\delta_{99}$ and the Zagarola and Smits velocity scale $u_{ZS} = u_e \delta_1 / \delta_{99}$. Results for part of their Fig. 3 are displayed here as our Fig. 3a. In Fig. 3b we plot the same velocity data using the *y*-axis scale $u/u_{ZS}$ instead of $(u_e - u)/u_{ZS}$. It is evident that the nine Smith velocity profiles scaled with $\delta_{99}$ and $u_{ZS}$ do not result in similar-like behavior in the outer region of the boundary layer.

The last example consists of another flow profile dataset used by Zagarola and Smits [2] in their paper. In Fig. 4a we reproduce a part of Zagarola and Smits Fig. 3 consisting of profiles taken from Purtell, Klebanoff, and Buckley [16]. The data shown in Fig. 4a collapse reasonably well in the outer region. Now consider the same velocity profile data plotted in Fig. 4b using the *y*-axis scale $u/u_{ZS}$ instead of $(u_e - u)/u_{ZS}$. Obviously, the five profiles in Fig. 4b do not show similar-like behavior in the outer region of the velocity boundary layer.

**5. Discussion**

In Figures 1-4, the results for the side-by-side comparisons are clear: whereas the defect plots appear to show similarity in the outer region of the boundary layer, the velocity profile plots do not. Hence the Zagarola and Smits velocity scale $u_{ZS}$ does not result in true similarity behavior for the datasets when cross-checked from their original papers [1-3]. Therefore, the Zagarola and Smits [1,2] velocity scale $u_{ZS}$ does not remove the effects of both the upstream conditions and finite local Reynolds number on the outer velocity profile as suggested by Castillo and George [4], rather it is normalizing the area under the scaled velocity profiles as we showed in Section 2. This normalization tends to make the defect profiles appear to be similar as long as the shape of the profiles is not significantly changing from one measurement station to the next. However, having equal areas is a necessary but not sufficient condition for similarity [14]. To have similarity it is still necessary to insure all of the requirements obtained by the flow governing equation approach to similarity [4,10-12] are satisfied. The results in Figs. 1-4 indicate that these profiles are not similar since not all of the requirements are met. This same type of flawed similarity was found by Weyburne [9] in eleven datasets used by Castillo, George, and coworkers [4-6] and three more datasets used by Panton [7] and Buschmann and Gad-el-Hak [8] to assert similarity using the Zagarola and Smits velocity scale. Hence, a careful reassessment of any claims in the literature is required for those groups that have attempted to use the defect profile and the Zagarola and Smits velocity scale.

The crucial point is that defect profile similarity can come about in two forms. True similarity occurs when both scaled defect profile similarity and velocity profile similarity is present. We follow Weyburne [9] and use the term “true similarity” in the sense that in this scenario, all the relevant constraints for similarity according to the flow governing equation approach to similarity [4,10-12] appear to be satisfied. The second form of defect similarity we term “false similarity” in the sense that although defect profile similarity is present, not all of the similarity criteria have been satisfied. In particular, the constraint $u_e/u_s$ be constant, is not satisfied. To ensure that one does not have this false form of defect profile similarity one can either generate plots of the $u_e/u_s$ and show that it is constant for the profiles in question. Alternatively, one can generate plots of the defect profiles and plots of the velocity profile and verify that similarity is present in both cases.

It may be tempting to try to dismiss the similarity equivalence argument offered above by claiming that the whole profile similarity as defined by Schlichting [13] is too restrictive, as it requires similarity in both the inner and outer regions. Note that in Eqs. 1 and 4 we assumed the equivalence holds for all *y*. Castillo and George’s [4] flow governing equation development was intentionally limited to the outer region of the boundary layer even though Castillo and George never defined exactly where the outer region boundary ends and the inner region begins. In the same spirit; examination of Eqs. 1 and 4 indicates that the only part of the equivalence argument that changes by restricting the argument to just the outer region is that instead of applying "for all y" in the general case it becomes "for all y in the outer region" for the case restricted to the outer boundary layer region. The requirement that $u_e/u_s$ be constant therefore still applies based on Castillo and George's [4] own similarity development for the outer region (see their Eq. 9) which means that defect profile similarity must be accompanied by velocity profile similarity even in the case where only the outer region is considered.

The true and false similarity argument rests on the fact that defect profile similarity must be accompanied by velocity profile similarity as shown in Section 3 above. To understand how this was not acknowledged earlier, we look back at Clauser's [17] early work on scaling of the turbulent boundary layer. Clauser's Figs. 2 and 3 compare a number of experimental datasets from various groups plotted as both velocity profiles and defect profiles. Whereas the defect profiles from different groups collapsed to a single universal profile, the velocity profiles did not. Subsequent searches for similarity scaling parameters for the turbulent boundary layer have adapted the use of the defect profile as a means of "discovering" similar behavior. Following closely after Clauser's work, Rotta [10] and Townsend [11] developed defect profile based theoretical treatments for the study of turbulent boundary layer similarity. This association of the defect profile with the turbulent boundary layer has been reinforced by the extensive work that has occurred on the turbulent boundary layers' inner region. The foundation for the Logarithmic Law of the Wall that describes the velocity profile behavior in the near wall region is the von Kármán's [18] analytical expression based on the defect profile. Given Clauser's early observation and the lack of any theoretical evidence to the contrary, it has been the Clauser view that the defect profile is "discovering" similarity behavior that the velocity profile did not find that has been the accepted paradigm. Hence, even though Rotta [10] and Townsend [11] knew that $u_e/u_s$ must be a constant, neither they, nor anyone else until Weyburne [9] made the connection that this condition necessarily also requires defect and velocity profile similarity must occur simultaneously for similarity to be present in a set of velocity profiles.

The condition that $u_e/u_s$ must be a constant was first theoretically derived by Rotta [10] (see his Eq. 14.3) and Townsend [11] (see his Eq. 7.2.3) using flow governing equation approaches to similarity. More recently, Castillo and George [4] (see their Eq. 9) and Jones, Nickels, and Marusic [12] (see their Eq. 3.9, a2+a4) also derived this criterion using the same type of flow governing equation approach. What all four flow governing equation approaches have in common is that their theoretical developments start off the same way as the flow governing equation approach used by many of the similarity treatments that have been successfully developed for various laminar flow boundary layer cases. Both the turbulent and laminar approaches start out by assuming that the *x* and *y* velocity components can be described as a product of an *x*-functional and a scaled *y*-functional. In each case these product functionals are then substituted into the conservation of mass and the Prandtl's boundary layer *x*-momentum balance equation. Similarity requires that each term in the transformed equation must change proportionately as one moves downstream (*x*-direction). Equivalently, if one divides through by one of the resulting *x*-groupings from the transformed equation, the requirement now becomes that the *x*-groupings ratios must be constants. The similarity flow conditions like $u_e/u_s$ equal to a constant are obtained at this step. The reason for this short summary is to point out that the $u_e/u_s$ equal to a constant constraint is based on the same steps and approach that has led to many of the laminar flow similarity results. Hence any attempt to dismiss the constraint will need to show not only which of the preceding steps has failed but also why it fails for the turbulent case but works for the laminar flow case. Furthermore, we offer three additional derivations of this criterion in Appendix A. The derivations in Appendix A are based on the definition of similarity (Eqs. 1 and 4) rather than the flow governing equation approach. Given the two separate theoretical derivation paths, it is unlikely that this criterion is incorrect.

Although there are differences in the theoretical approaches, all the flow governing equation formulations have as a similarity requirement that $u_e/u_s$ is constant (or equivalent). Recent scaling reviews [7,8] insist that it is the Castillo and George formulation that requires $u_e$ must be an outer layer velocity scaling parameter. In fact, Rotta and Townsend came up with the same criterion more than forty years previously. The turbulent boundary layer scaling review papers [7,8] go on to argue that the flow governing equation approach to similarity developed by Castillo and George is problematic since $u_e$ makes a poor outer region scaling parameter compared to $u_{ZS}$ or $u_\tau$, the friction velocity judging by comparing experimental plots. The reviews offered a number of experimental data plot comparisons as proof. Weyburne [9] examined the experimental data offered as proof of this assertion and showed that in each case the reviewers used defect profile plots coupled with $u_{ZS}$ or $u_\tau$ that turned out to be cases of false similarity. For example, both Panton [7], and Buschmann and Gad-el-Hak [8] used some of the Österlund's [19] data to assert that $u_e$ makes a poor outer region scaling parameter compared to $u_{ZS}$ or $u_\tau$. In Fig. 5 we plot some of Österlund's [19] data as scaled defect and scaled velocity profiles using $u_{ZS}$ as the scaling parameter. The side-by-side comparison indicates that this dataset does not display true similarity behavior. Hence any comparisons between $u_{ZS}$ and $u_e$ are meaningless for this dataset since the dataset does not show true similarity when scaled with $u_{ZS}$. The same was found for the other datasets used by Panton [7], and Buschmann and Gad-el-Hak [8]. Hence, their attempt to discredit $u_e$ as a valid similarity scaling parameter, and by inference the flow governing approach to similarity, must be rethought.

To further counter the perception that there is a problem with the flow governing equation approach because they all point to $u_e$ as a similarity scaling parameter, consider some experimental evidence in Figs. 6-11. In these figures the velocity scale is $u_e$ and the length scale is the displacement thickness $\delta_1$. The first of these figures, Fig. 6, shows an example of similar-like behavior using $u_e$ as the scaling variable. Note that Fig. 6 is a subset of the Smith [15] data from Fig. 3 that showed false similarity behavior when scaled with $u_{ZS}$. Another example of similar-like behavior using $u_e$ is shown in Fig. 7. This is the same Österlund [19] data from Fig. 5 that proved to be false similarity behavior when scaled with $u_{ZS}$. Choosing $u_s = u_e$ means that the Eq. 5 criterion will automatically be satisfied so that the behavior in Figs. 6-7 is expected for any dataset displaying similarity. The next four figures, Figs. 8-11, are based on experimental data from Ludwieg and Tillmann [20], Clauser [17], Jones, Marusic, and Perry [21], and Sillero, Jiménez, and Moser [23]. These figures demonstrate: 1) that a number of datasets displaying similar-like behavior do exist, 2) that $u_e$ is a legitimate velocity scale for similarity, and 3) that there are examples where $u_e$ appears to be superior to $u_{ZS}$ in terms of producing similar behavior.

It should be noted that Weyburne [14] recently used rigorous mathematics to prove that if similarity exists in a set of 2-D boundary layer profiles, then the similar length scale $\delta_s$ must be proportional to $\delta_1$ and that the similar velocity scale $u_s$ must be proportional to $u_e$. This combination of scaling parameters also insures that the $a(x)=1$ in Eq. 2, just as was the case for the Zagarola and Smits scaling parameters. The difference is that the Weyburne scaling automatically satisfies the flow governing equation requirement that $u_e/u_s$ must be a constant. Comparisons of the two scaling parameter sets indicate that it is the boundary layer thickness $\delta$ assumption that is the problem in the Zagarola and Smits scalings discussed above.

In spite of the evidence presented above, it is possible that there is an alternative explanation that fits the data equally well. However, such an alternative explanation would need to address two issues: 1) Those who claim that $u_{ZS}$ is superior to $u_e$ as an outer region scaling parameter need to explain the results in Figs. 6-11, and 2) How can one explain defect profile similarity Figs. 1a-4a but not velocity profile similarity in Figs. 1b-4b. At the present time, the results presented in Figs. 1-4 cannot be explained other than with the true and false similarity explanation given above.

## 6. Conclusions

Based on the above analysis, we conclude that claims of similarity based on the defect profile need to ensure that all criteria for similarity have been satisfied. For the datasets used by the Zagarola and Smits [1,2] and McKeon and Morrison [3], none of the claims for similarity included a check to ensure that $u_e/u_s$ is a constant. When compliance was checked herein, it was found that although the datasets displayed defect similarity, they did not satisfy the criteria that $u_e(x)/u_{ZS}(x)$ must be constant for similarity. Hence these datasets do not display true similarity. Given that this same similarity problem was identified in Castillo, George and coworkers [4-6] work using the Zagarola and Smits scaling parameter, a careful reassessment of claims is indicated for those groups who use the defect profile and the Zagarola and Smits velocity scaling parameter to assert similarity of the velocity profile of turbulent boundary layers.

## Acknowledgement

The author acknowledges the support of the Air Force Research Laboratory and Gernot Pomrenke at AFOSR. In addition, the author thanks the various experimentalists for making their datasets available for inclusion in this manuscript.

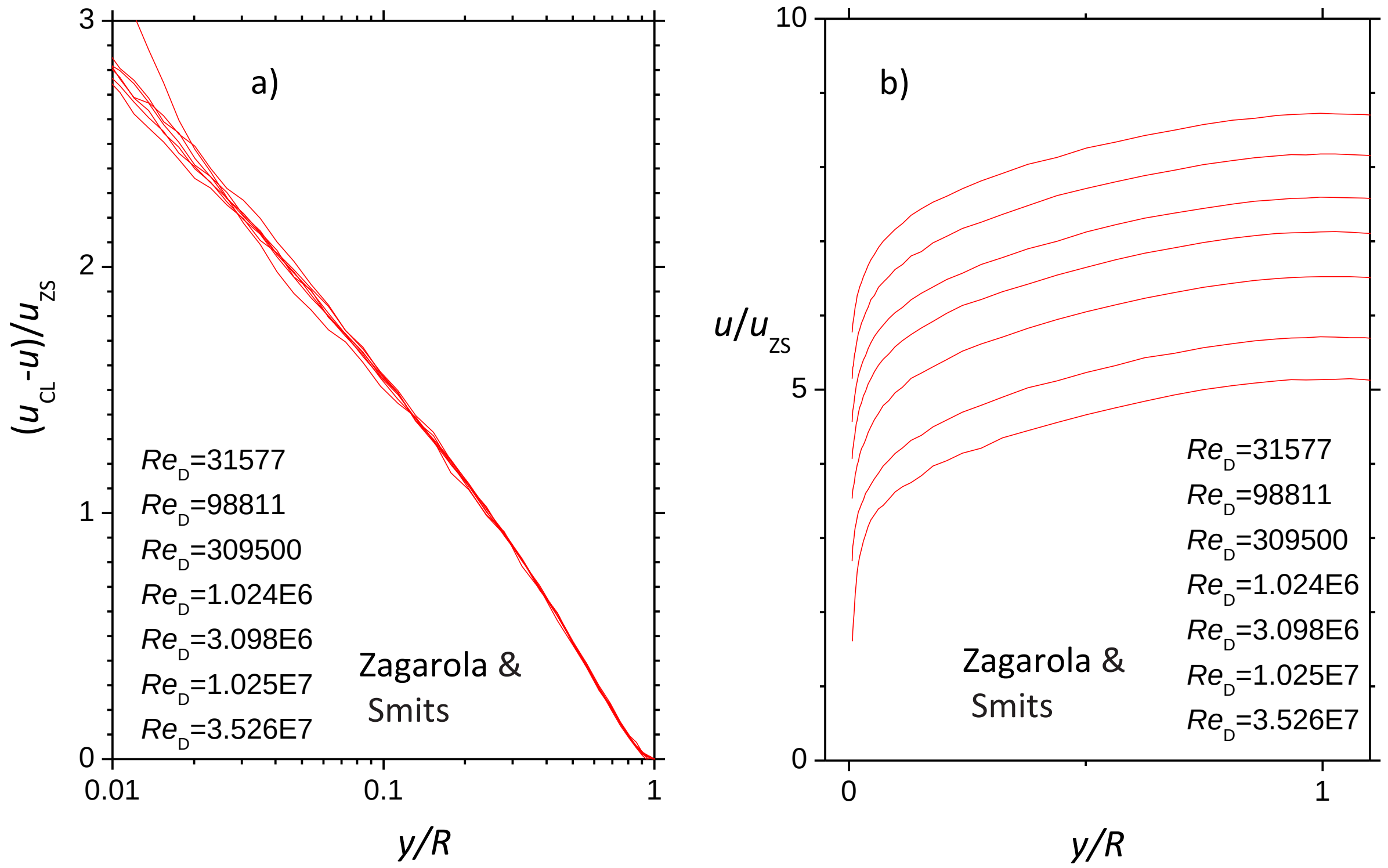

Fig. 1: Zagarola and Smits' [1] pipe data plotted as a) defect profiles and b) velocity profiles using $u_s = u_{ZS}$. The $Re_\theta$ list identifies which profiles are plotted.

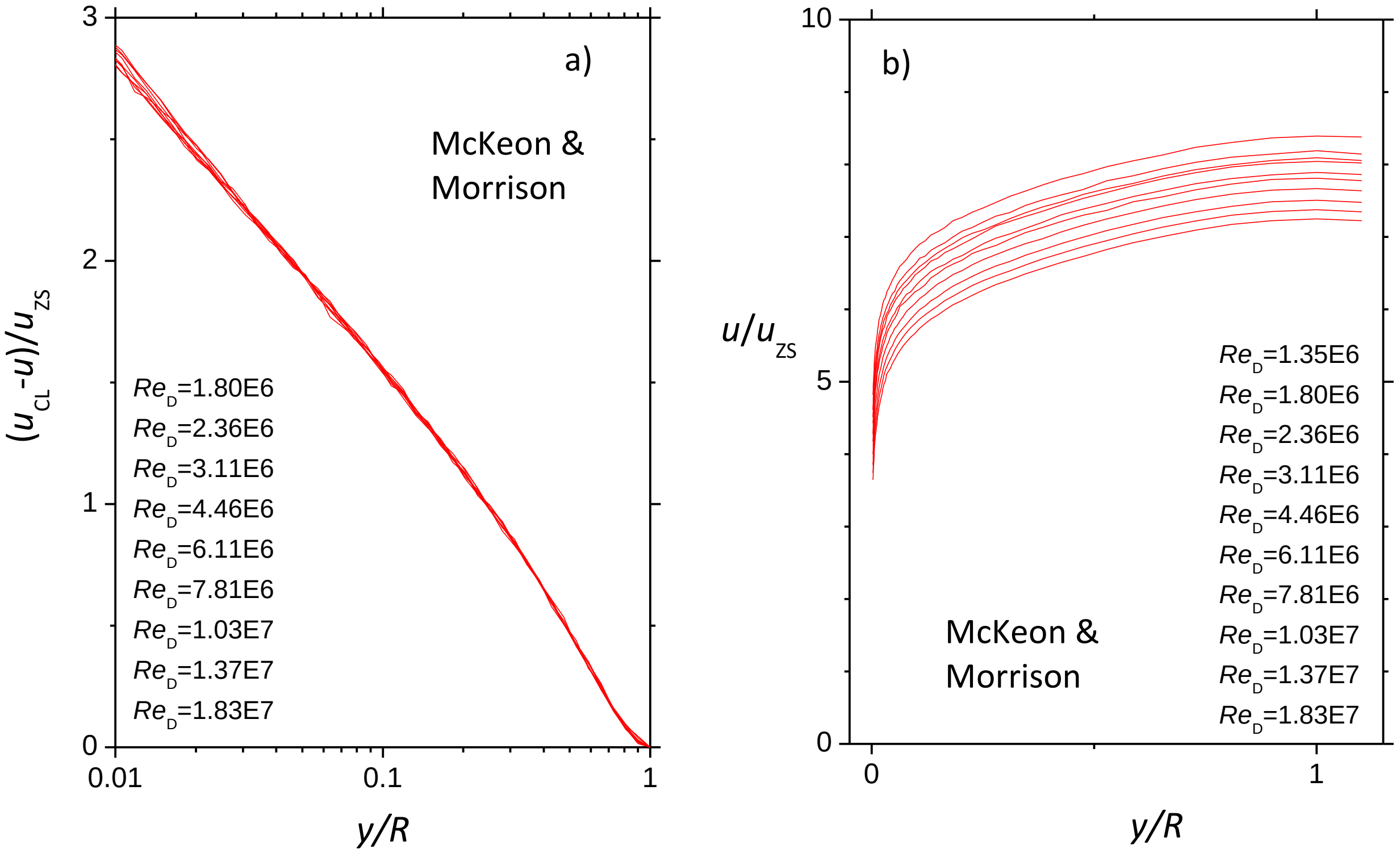

Fig. 2: McKeon and Morrison's [3] pipe data plotted as a) defect profiles and b) velocity profiles using $u_s = u_{ZS}$. The $Re_\theta$ list identifies which profiles are plotted.

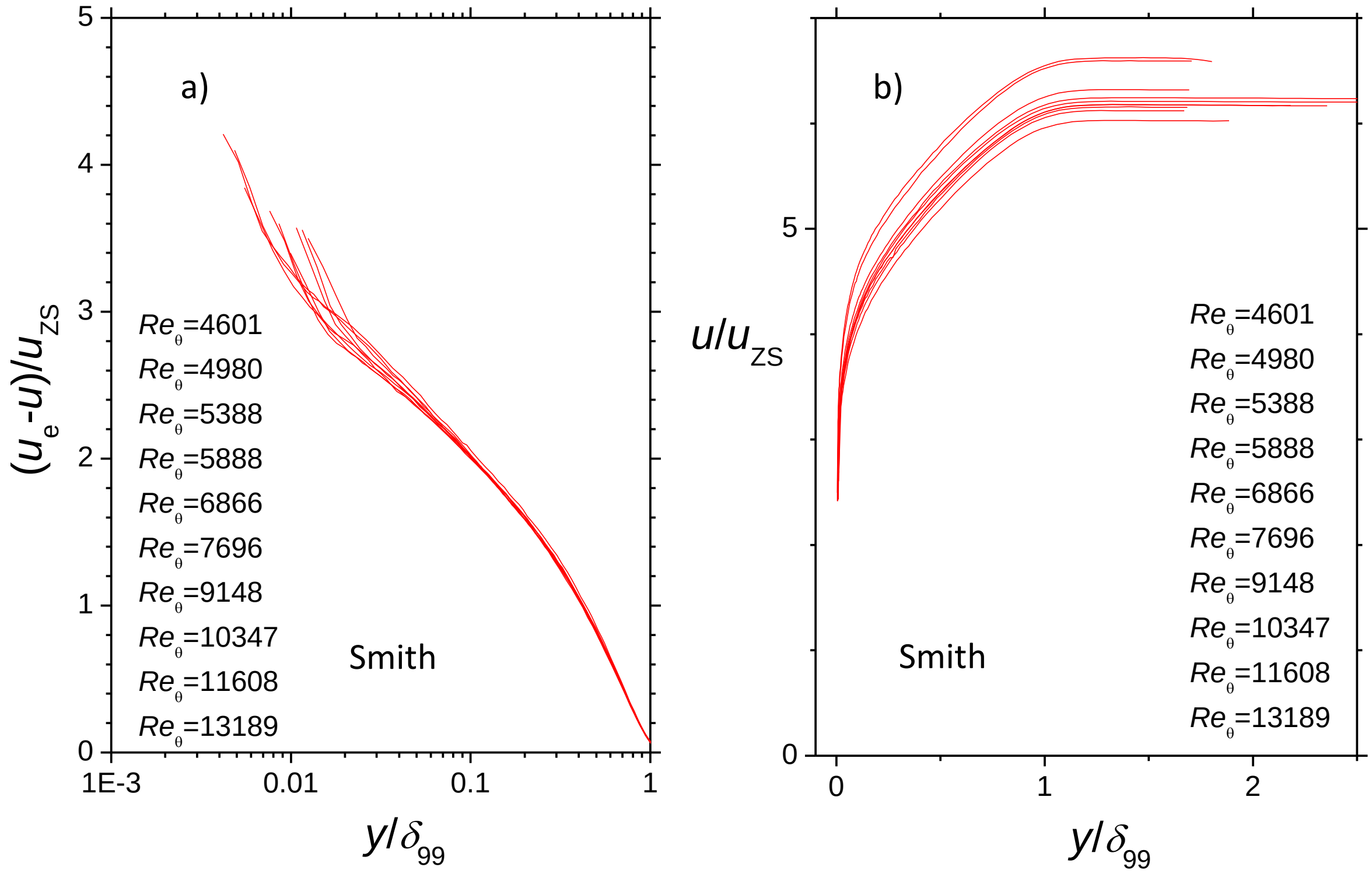

Fig. 3: Smith's [15] data plotted as a) defect profiles and b) velocity profiles using $u_s = u_{ZS}$. The $Re_\theta$ list identifies which profiles are plotted.

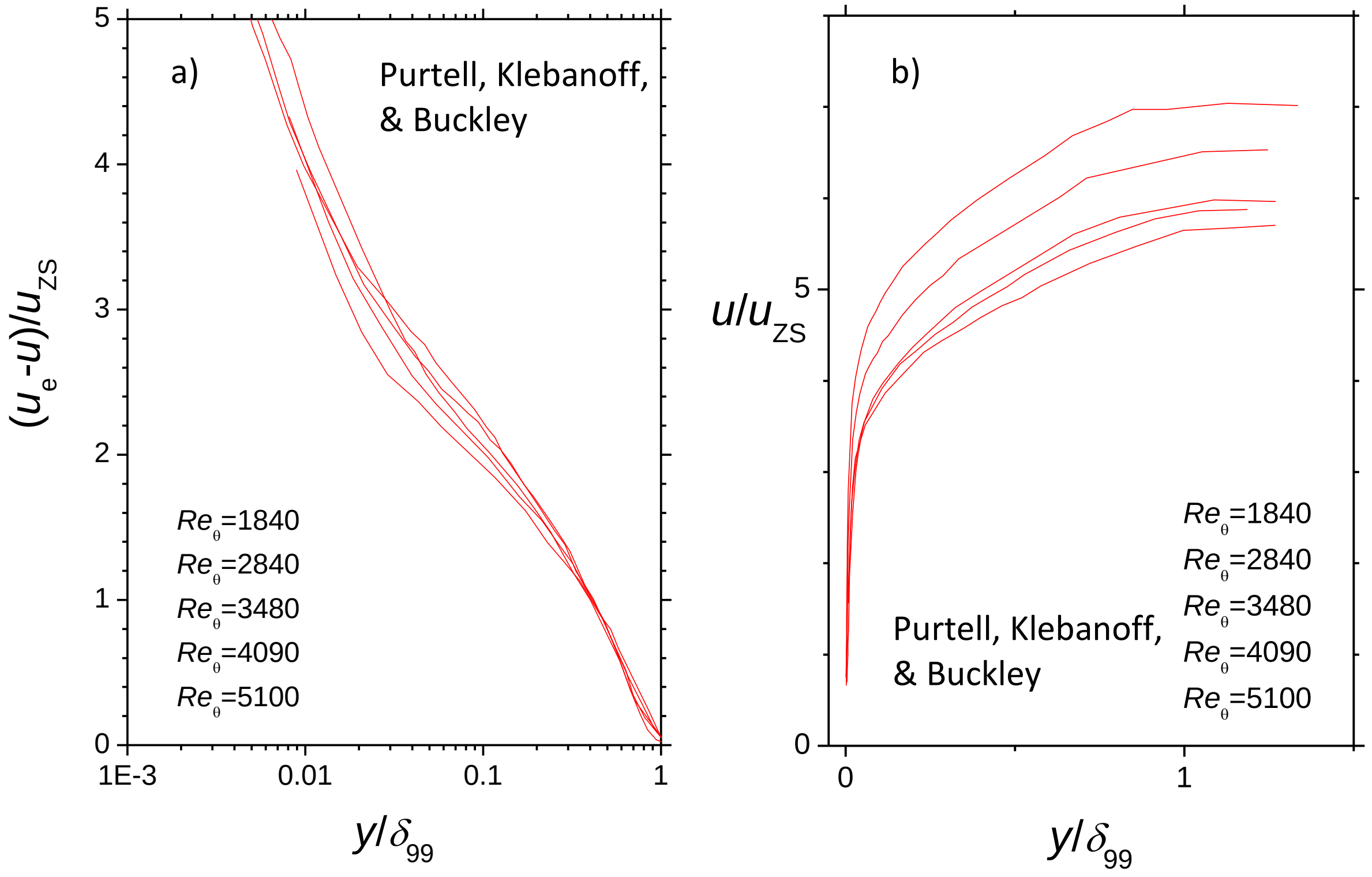

Fig. 4: Purtell, Klebanoff, and Buckley's [16] data plotted as a) defect profiles and b) velocity profiles using $u_s = u_{ZS}$. The $Re_\theta$ list identifies which profiles are plotted.

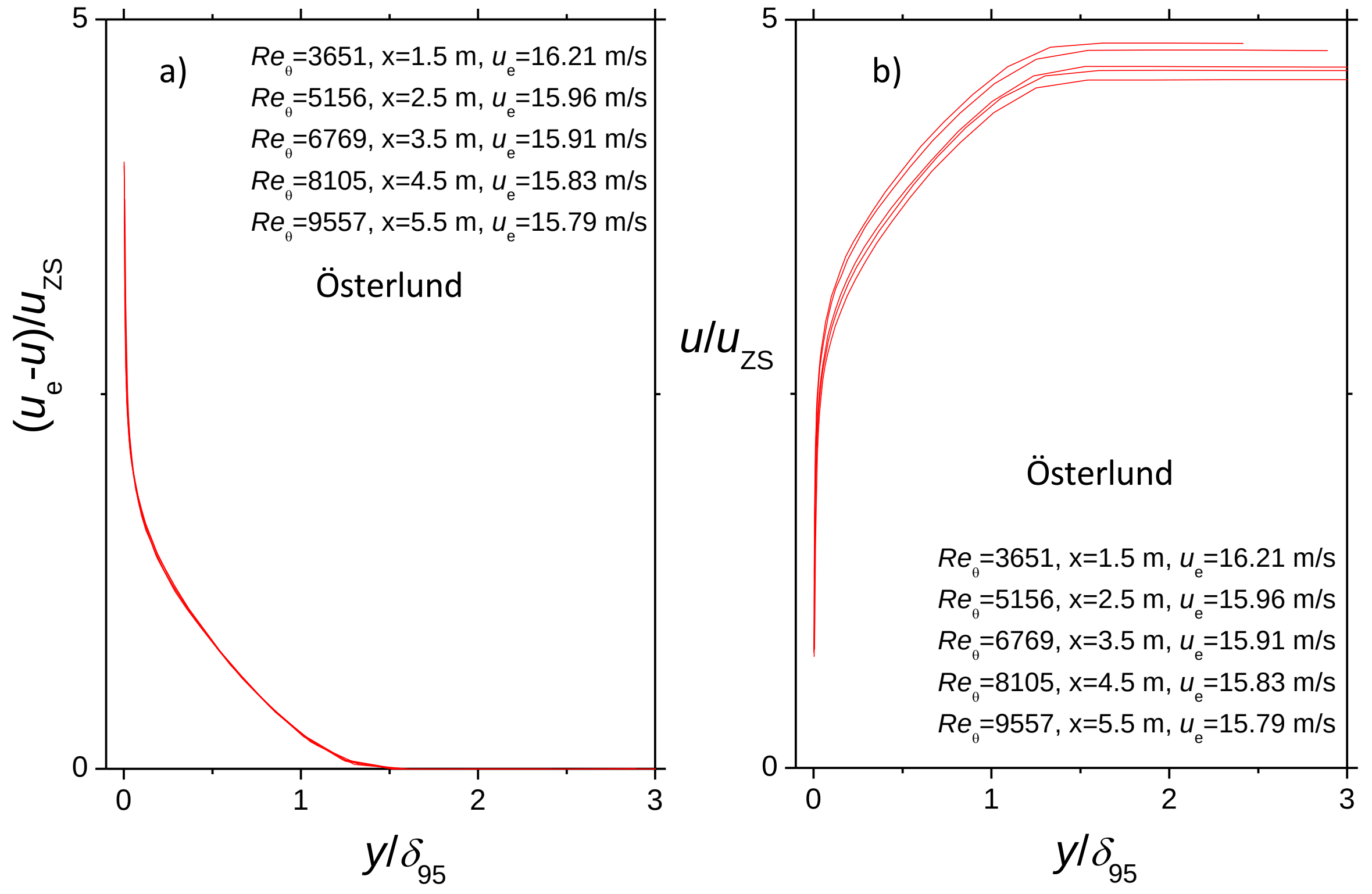

Fig. 5: Some of Österlund's [19] data plotted as a) defect profiles and b) velocity profiles using $u_s = u_{ZS}$. The $Re_\theta$ list identifies which profiles are plotted.

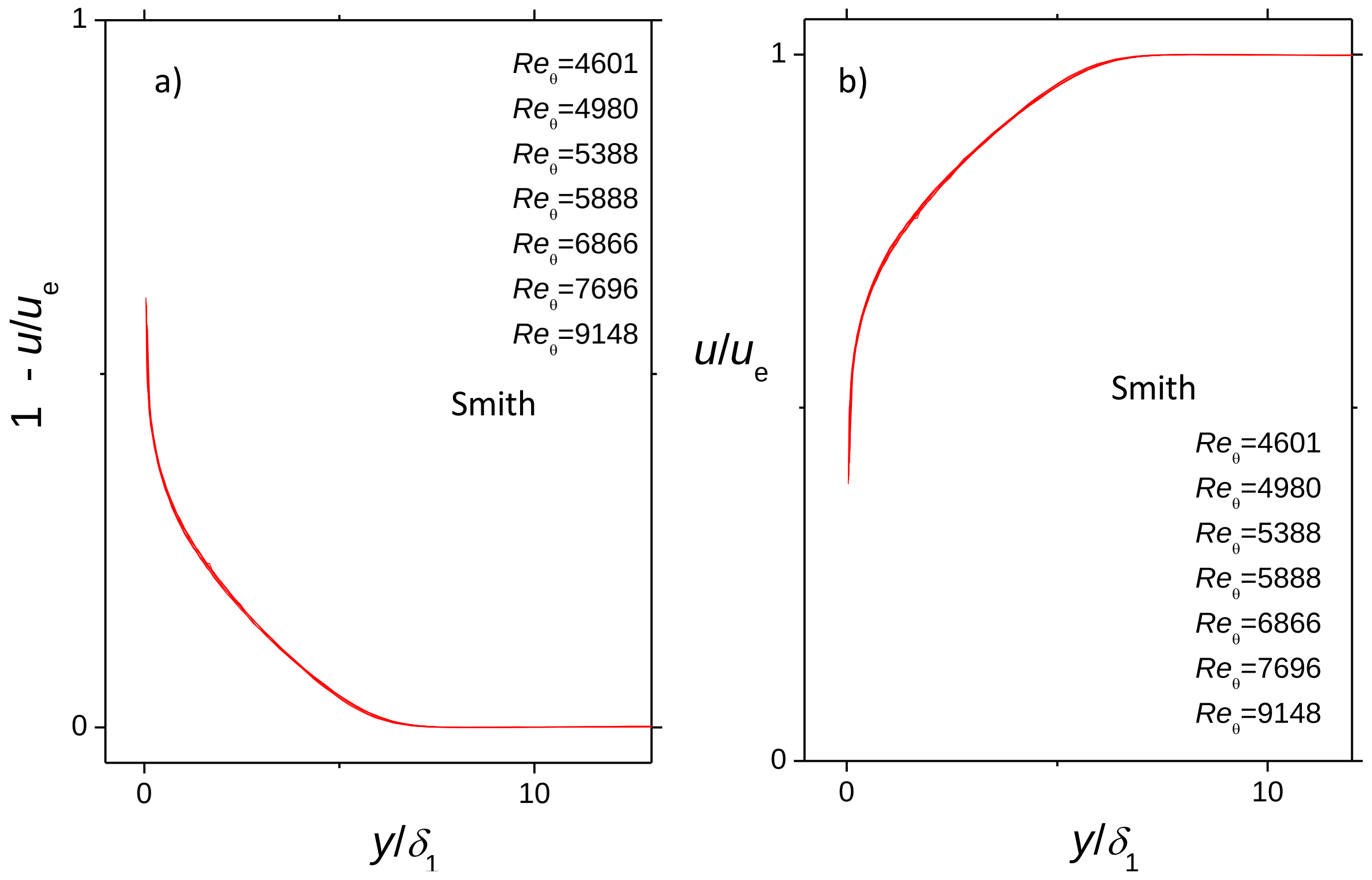

Fig. 6: Smith's [15] data plotted a) as defect profiles and b) as velocity profiles using $u_s = u_e$. The $Re_\theta$ list identifies which profiles are plotted. Compare these plots to those in Fig. 3.

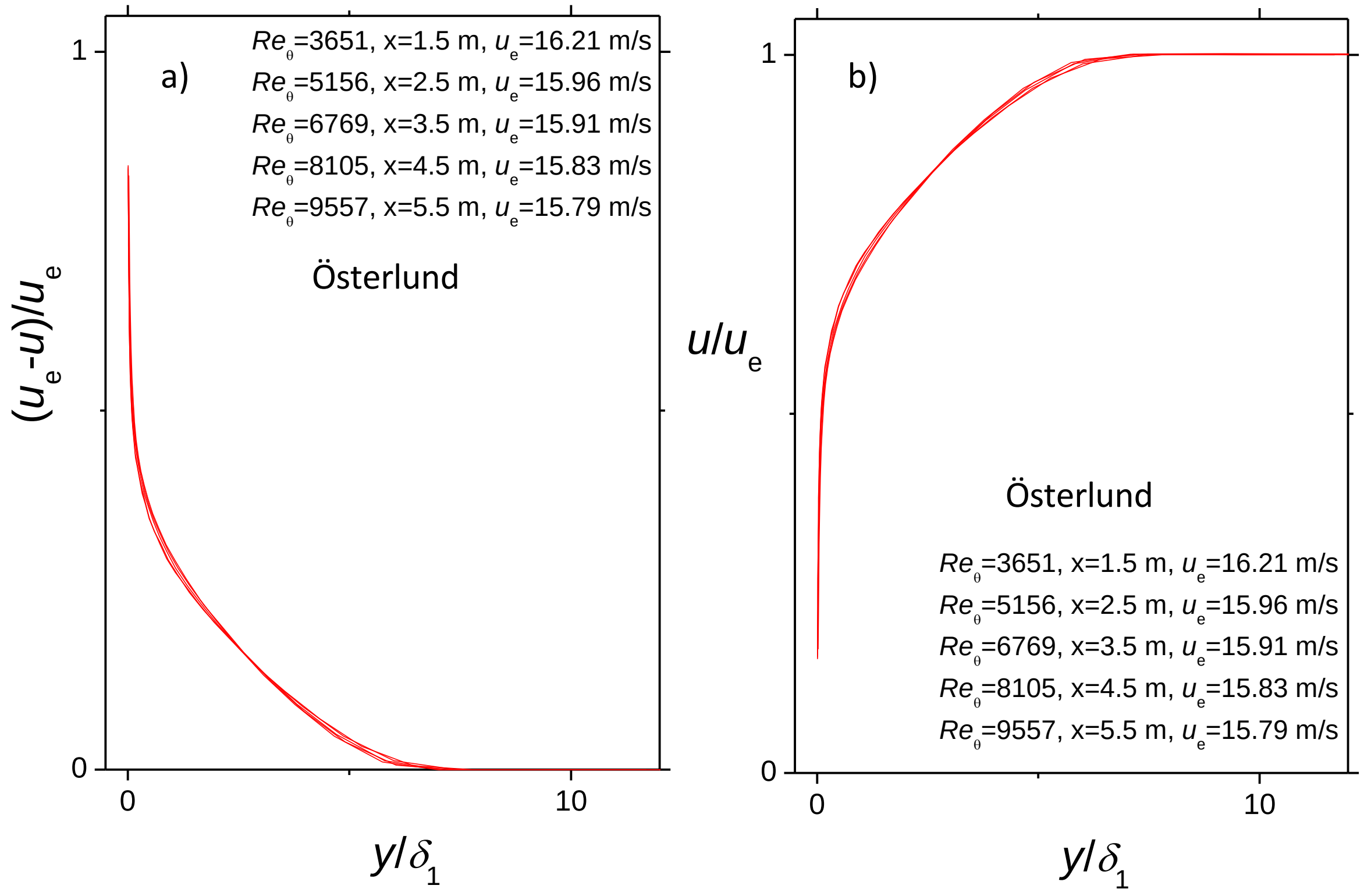

Fig. 7: Österlund's [19] data plotted a) as defect profiles and b) as velocity profiles using $u_s = u_e$. The $Re_\theta$ list identifies which profiles are plotted. Compare these plots to those in Fig. 5.

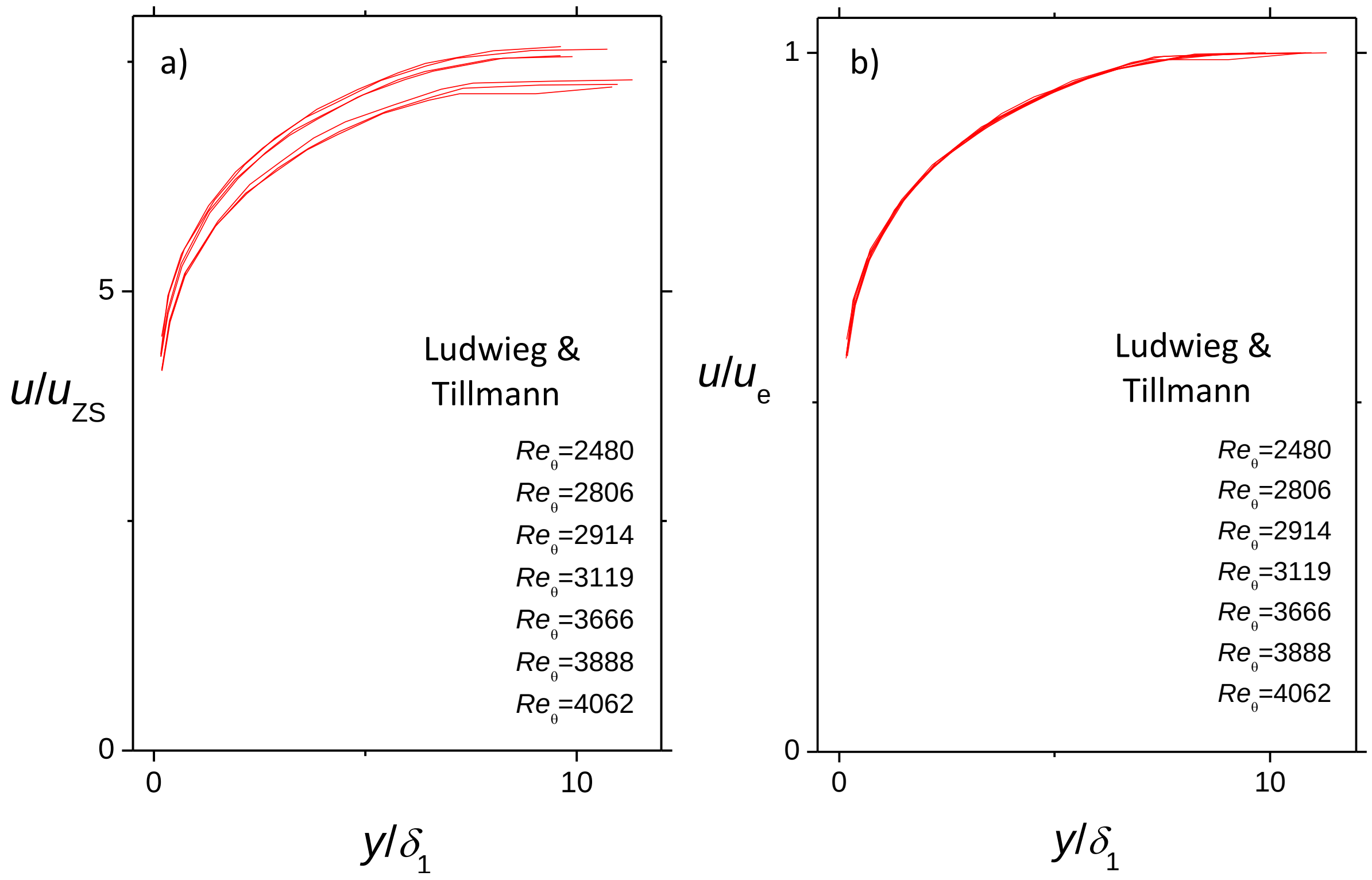

Fig. 8: Ludwieg and Tillmann's [20] data plotted a) using $u_s = u_{ZS}$ and b) using $u_s = u_e$. The $Re_\theta$ list identifies which profiles are plotted.

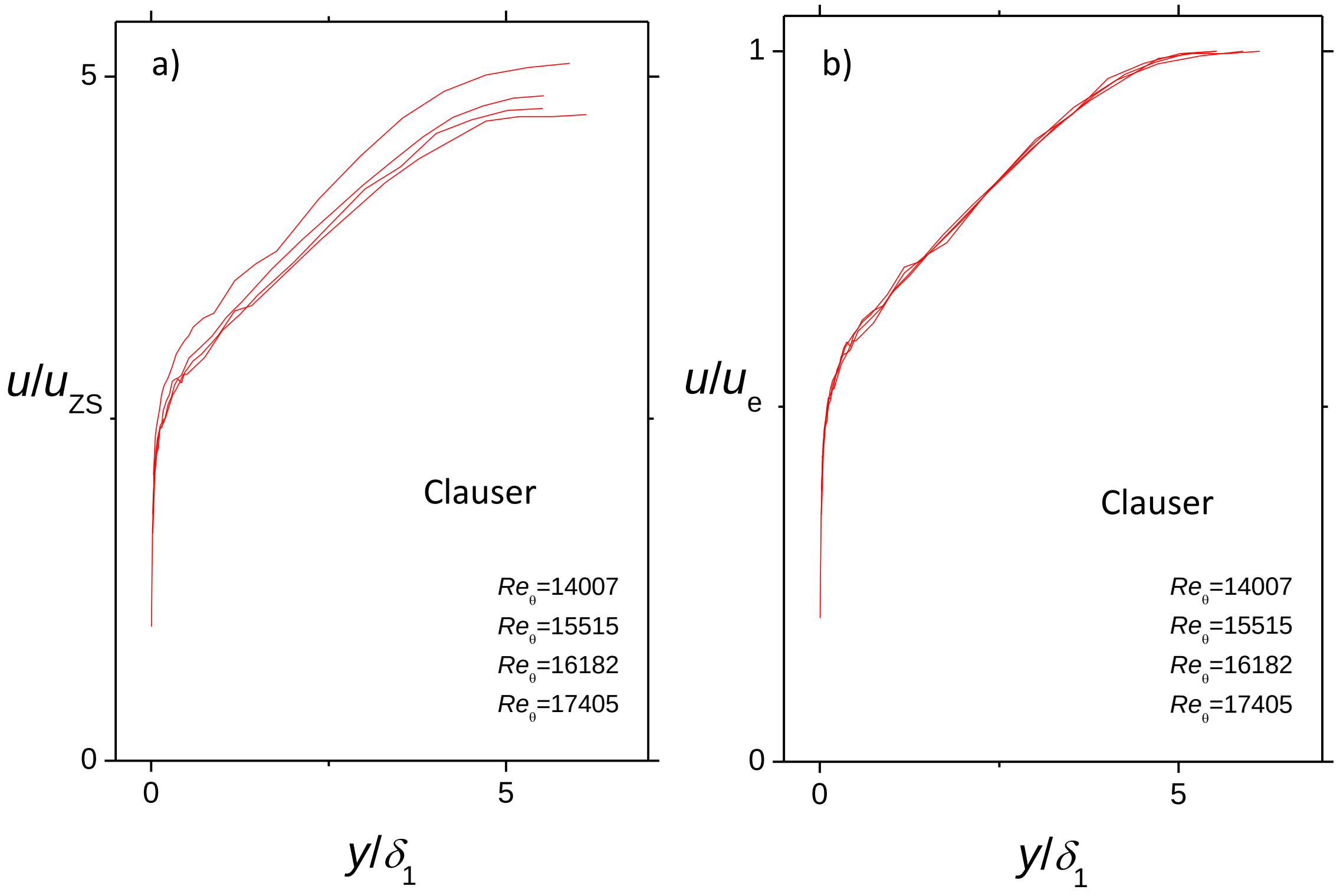

Fig. 9: Clauser's [17] data plotted a) using $u_s = u_{ZS}$ and b) using $u_s = u_e$. The $Re_\theta$ list identifies which profiles are plotted.

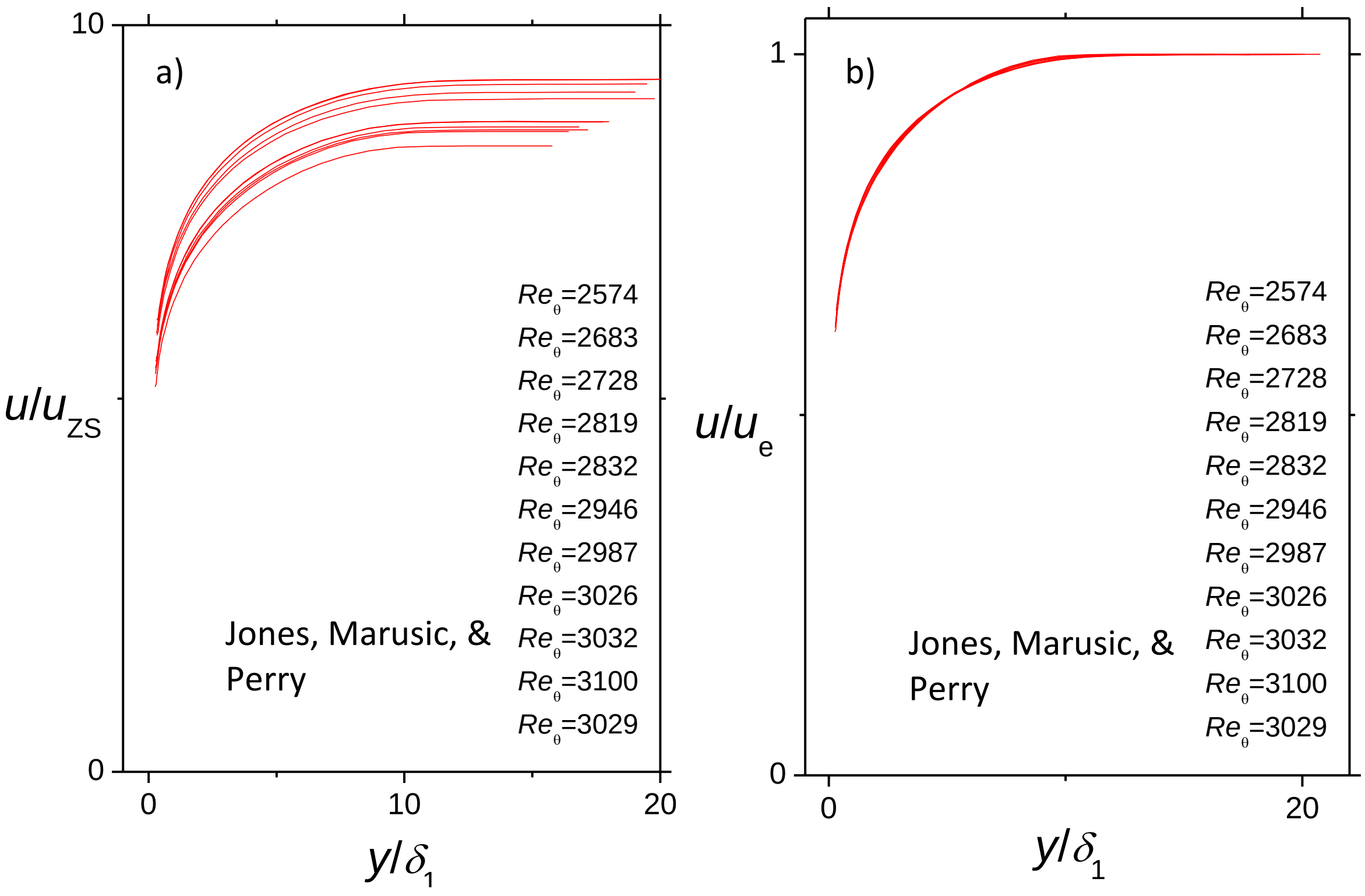

Fig. 10: Jones, Marusic, and Perry's [21] data plotted a) using $u_s = u_{ZS}$ and b) using $u_s = u_e$. The $Re_\theta$ list identifies which profiles from the K = 2.70 × $10^{-7}$ dataset are plotted.

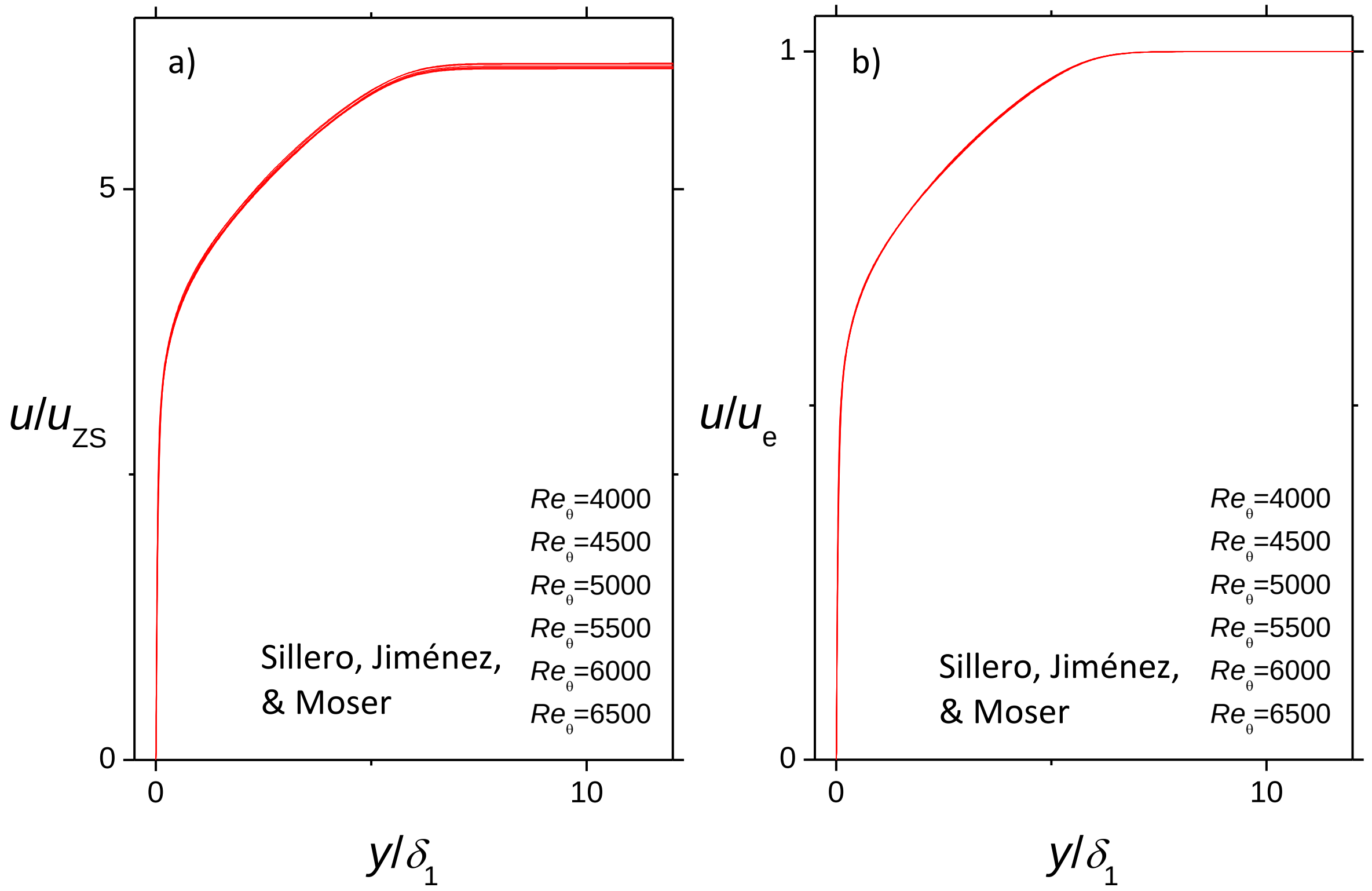

Fig. 11: Sillero, Jiménez, and Moser's [23] DNS data plotted a) using $u_s = u_{ZS}$ and b) using $u_s = u_e$. The $Re_\theta$ list identifies which profiles are plotted.

## Appendix A: Alternative Derivations of the Velocity Ratio Scaling Criterion

The flow governing approaches to similarity by Rotta [10] (see his Eq. 14.3), Townsend [11] (see his Eq. 7.2.3), and Castillo and George [4] (see their Eq. 9) and Jones, Nickels, and Marusic [12] (see their Eq. 3.9, a2+a4) all have as a similarity requirement that $u_e / u_s$ is equal to a constant, (or equivalent). The flow governing approach is not the only theoretical route to this criterion. We can offer at least three additional simple theoretical routes to substantiate $u_e$ as a valid similarity parameter choice.

First, start with the definition of defect similarity given by

$$\frac{u_e(x) - u(x, y/\delta_s)}{u_s(x)} = f(y/\delta_s) \quad , \tag{A.1}$$

where *f* is the universal profile function of the dimensionless height. Note that "for all *y*" is assumed to be the general case. Since this equation must hold for all *y*; take the case where *y*=0 so that Eq. A.1 becomes

$$\frac{u_e(x)}{u_s(x)} = f(0) \quad , \tag{A.2}$$

where $f(0)$ is a constant. Therefore, $u_e$ must be a valid similarity parameter choice.

Now let us consider a second theoretical route. Weyburne [14] recently used rigorous mathematics to prove that if similarity exists in a set of boundary layer profiles, then the displacement thickness $\delta_1$ must be a similar length scale and $u_e$ must be a similar velocity scale. This derivation starts with the traditional definition of similarity given according to Schlichting [13] as

$$\frac{u(x_1, y/\delta_s(x_1))}{u_s(x_1)} = \frac{u(x_2, y/\delta_s(x_2))}{u_s(x_2)} \quad . \tag{A.3}$$

If similarity is present in a set of velocity profiles then it is self-evident that the properly scaled first derivative profile curves (derivative with respect to the scaled *y*-coordinate) must also be similar. It is also self-evident that the areas under the scaled first derivative profiles plotted against the scaled *y*-coordinate must be equal for similarity. In mathematical terms, the area under the scaled first derivative profile curve is expressed by

$$a(x) = \int_0^{h/\delta_s} d\left\{\frac{y}{\delta_s}\right\} \frac{d\left\{u(x, y/\delta_s)/u_s\right\}}{d\left\{\frac{y}{\delta_s}\right\}} \quad , \tag{A.4}$$

where $a(x)$ is in general a non-zero numerical value and $y = h$ is located deep in the free stream. Using a variable switch $\left(d\{y/\delta_s\} \Rightarrow (1/\delta_s)dy\right)$, Eq. A.4 can be shown to reduce to

$$a(x) = \frac{u_e(x)}{u_s(x)} \quad . \tag{A.5}$$

Similarity requires that $a(x_1) = a(x_2) = \text{constant}$. Once again, $u_e$ must be a valid similarity parameter choice.

A third theoretical route also starts with Eq. A.3. Since this equation must hold for all *y* (or at least all y in the outer region), then one can take the case where $y \to h(x)$, $h(x)$ is located deep in the free stream. This means that $u(x, h(x)) = u_e(x)$. Similarity assumes that we have chosen the same scaled *y*-value so that we must have

$$\frac{h(x_1)}{\delta_s(x_1)} = \frac{h(x_2)}{\delta_s(x_2)} \quad . \tag{A.6}$$

This is easily satisfied since we are free to choose $h(x_1)$ and $h(x_2)$ to be located anywhere in the boundary layer edge region. If we choose the *y*-values $h(x_1)$ and $h(x_2)$ to be above the boundary layer edge, then Eq. A.3 reduces to

$$\frac{u_e(x_1)}{u_s(x_1)} = \frac{u_e(x_2)}{u_s(x_2)} \quad . \tag{A.7}$$

While other velocity scaling parameters are not excluded, we have presented three related theoretical formulations, all of which indicate that if similarity exists in a set of velocity profiles then the velocity at the boundary layer edge $u_e$ must be a similarity scaling parameter for the 2-D boundary layer. Notice that none of these derivations in this Section are based on the flow governing equations but rather on the definition of similarity itself. Hence these arguments apply to all 2-D boundary layer flows which show similarity from station to station.